\documentstyle[BIB]{report}    % Specifies the document style.
\newcommand{\be}{\begin{equation}}
\newcommand{\ee}{\end{equation}}
\def\lsim{\raise0.3ex\hbox{$<$\kern-0.75em\raise-1.1ex\hbox{$\sim$}}}
\def\gsim{\raise0.3ex\hbox{$>$\kern-0.75em\raise-1.1ex\hbox{$\sim$}}}

\title{Multi-quark energies in QCD.}
\author{A.M. Green\thanks{Bitnet address:GREEN@FINUHCB},\\
 Research Institute for Theoretical Physics,  University
of Helsinki, Finland, \\
 C. Michael\thanks{E-mail address:CM@UKACRL},\\
DAMTP, University of Liverpool, Liverpool L69 3BX, UK \\
 and J.E. Paton\thanks{E-mail address:PATON@PH.OX.AC.UK},\\
Department of Physics: Theoretical Physics, University of Oxford, UK.}

\begin{document}

\large

\maketitle

\begin{abstract}

Four-quark potentials for $SU(2)$ are evaluated in the static limit
with the quenched approximation -- using a lattice of $16^3\times 32$
and $\beta=2.4$. The four quarks are restricted to the corners of rectangles
with sides upto seven lattice spacings long. The results are analysed in terms
of a strategy based on interquark two-body potentials -- as advocated earlier
by the authors. This shows that a standard two-body approach overestimates
the four-quark binding energy by upto a factor of about three for the largest
rectangles.
\end{abstract}
\newpage

\section{ Introduction}
\setcounter{equation}{0}

\vskip 0.5 cm

The words -- "Quantum Chromodynamics (QCD) is the theory
describing the strong interactions between elementary particles"
-- often form the opening sentence of articles discussing the interactions
between quarks
and gluons. Unfortunately, this expectation is far from being completely
proven theoretically. Only in those extreme situations involving short
distances or high momenta, where perturbation theory is applicable,
has the theory been satisfactorily verified. In comparison, at more
typical interquark distances of $\approx\frac{1}{2}$ fm -- those found in
mesons and baryons -- the consequences of this
theory can not be evaluated in any simple direct
manner. This has led to the so called Monte Carlo lattice simulations,
where the basic equations of QCD are expressed in terms of a discretised
space-time. For this to be realistic, when describing a hadron of size
$\approx 1$ fm, the space-time volume $L^3T$, in which the theory is being
tested, must be sufficiently large to avoid surface artefacts i.e. $L\approx
2$ fm. On the other hand, the lattice covering this volume must have
a spacing $a$ that is sufficiently small to be a meaningful simulation
of the original QCD problem in which $a$ should tend to zero. In practice,
this seems to require $a$ to be no larger than $\approx 0.1$ fm, which
 implies that each
spatial direction must be discretised into at least $n=\frac{L}{a}\approx 20$
steps.
A similar number of steps is also required in the time
direction\renewcommand{\thefootnote}{1}\footnote{\normalsize In practice it is
the
imaginary
time $T=it$ that is used in this formulation of QCD.} to ensure
an accurate evaluation of observables. The theory is then expressed in terms
of quantities on the links connecting the $L^3T$ lattice points. This
leads to the number of degrees of freedom, which -- because
of the nature of the Monte Carlo simulation -- needs to be continuously
updated, being in the region of $10^7$, at least. Since it is necessary
to deal many times with all of these degrees of freedom, the problem is very
heavily computer orientated. However,
in spite of the tremendous  amount of computer time that has been devoted to
the
above method of dealing with QCD, the problems actually considered are very
modest. These have been mainly in the so-called pure gluon phase, in which
there are no quarks, and in the single meson or baryon sectors involving
only 2 or 3 quarks. Virtually nothing reliable has been said in this approach
about
multi-quark systems, since the appropriate Monte Carlo simulations require
lattices that are too large to be treated by present day computers. This is
unfortunate, since much of particle and nuclear physics involves the
interactions between hadrons i.e. systems requiring more than  3 quarks
for their description.

At this point it is worth remembering how other interacting multiparticle
systems have been treated. One possible analogy to the present
situation of many quarks interacting via gluons is that of many nucleons
interacting via mesons. In the latter case, much work has been devoted
to first understanding and parametrizing the two-nucleon problem in terms
of an internucleon potential $V_{NN}(r)$, in which there is no explicit
appearance of the underlying meson degrees of freedom. It is then found
that many-nucleon systems are well described using simply this two-body
interaction -- the corrections due to explicit mesonic effects being, in
general, much less than $10\%$. In order to test whether or not a similar
two-body approach is accurate in the case of quarks and gluons, two
quantities are, therefore, required:

1) A reliable interquark potential $V_{qq}(r)$ -- often called simply
$v(r)$ later.

2) For some multi-quark state, an accurate energy with which to
compare the

two-body approach based on $V_{qq}(r)$.

\noindent For simplicity, both of these quantities will only be considered
for the case of infinitely heavy quarks. This means that only systems
involving the heaviest quarks, i.e.  the bottom or charmed quarks, are
appropriate  in any comparison with experiment. This is a major difference
in comparison with the analogous multi-nucleon description of nuclei,
where to start with the system is essentially non-relativistic.

Nowadays, the potential  between two heavy quarks
seems to be well determined and well
parametrized by the form
\be
V_{qq}(r)=-\frac{e}{r}+b_Sr
\ee
i.e. a short range Coulomb-like term and a long range term, which
confines the quarks. Both of these have a simple theoretical
interpretation and the actual constants $e,b_S$ can be extracted from
fitting the masses of heavy mesons, which are essentially bound states in this
potential. In addition, Monte Carlo lattice simulations for the two-quark
system are consistent with this $V_{qq}(r)$, as will be discussed in
section 2.

Since the two-quark system is part of the basic input giving $V_{qq}(r)$
and the three-quark system concerns only single baryons (e.g. see \cite{som}),
the smallest number of quarks for which the present discussion on hadron
interactions becomes meaningful is four.
Attempts have been made in the past \cite{Markum,Ohta} to evaluate the
energy of four-quark systems, but the results were very inaccurate -- the
calculation most relevant to the present work being made in 1985 by
 Ohta et al.\cite{Ohta} . In this, four quarks were placed at the corners
of a rectangle in a four-dimensional lattice of size $16^3\times 6$ and
$\beta=5.8$ -- corresponding to a spacing
of $a\approx 0.15$ fm. Unfortunately, as a function of the rectangle size,
the resulting energies had error bars that were too large to test whether
or not a two-body description with $V_{qq}(r)$ was accurate. One of the
purposes of the present work was, therefore, to redo the above
four-quark calculation, but with the improvements that have occurred
since 1985. These are basically twofold -- i) at least an order of magnitude
improvement in the power of computers and ii) developments\cite{CM1,CM2} in the
theory of Monte Carlo lattice simulations leading to orders of magnitude
improvement in efficiency. In addition, to gain yet another order of magnitude
in computer efficiency,
 the use of $SU(3)$ in ref.\cite{Ohta} is replaced by $SU(2)$. For the two
 body problem, it has been shown \cite{CM1,CM2} that $SU(2)$ is
perfectly good to test models and, in practice, it is quite
close to $SU(3)$. Refs.\cite{CM1,CM2} also show that, for the values of
$\beta$ being used, scaling -- but not asymptotic scaling -- has set in.
It is now hoped that, for the four quark potentials, $SU(2)$ leads to an
equally good approximation for the more realistic $SU(3)$ situation and
also that scaling has been achieved. It should be added that in $SU(2)$,
quarks and antiquarks are for most purposes essentially identical.
Therefore, from now on in the discussion
no reference will be made to antiquarks i.e. meson-like configurations
will be composed of two quarks.

As will be discussed in section 3, in the present work  a lattice of size
$16^3\times 32$ with a $\beta=2.4$ -- corresponding to a lattice spacing of
$a \approx 0.12$ fm -- is treated. The use of 32 steps in the $T$ direction,
compared with only 6 in ref.\cite{Ohta}, also means that the present system
is closer to the zero temperature limit -- a limit that is assumed in the
subsequent two-body approach of section 4.

As mentioned above, all four quark masses are assumed to be infinite
and also quark-antiquark  pair creation is neglected-- the so-called quenched
approximation. In the same way as refs.\cite{Markum,Ohta,GMP},
the actual configuration
treated is that shown in fig. 1a), where the four quarks are located at the
corners
of a rectangle with sides of length $r$ and $d$. However, in the subsequent
analysis
it is convenient to consider the total energy $(E_4)$ of this configuration
in terms of an interaction involving the two meson-like partitions
$A=M_{13}M_{24}$ and $B=M_{14}M_{23}$ in figs. 1b) and 1c), where, for example,
$M_{13}=[q_1q_3]^0$ -- the superfix being the overall colour index.
The four-quark potential energy $V_4(d,r)$ is then taken to be
\be
\label{V4}
V_4(d,r)=E_4(d,r)-2V_{qq}(d)
\ee
Here the $V_{qq}(d)$ can be thought of as the internal energy of one of the
two mesons $M_{13}$ and $M_{24}$, which are interacting through a potential
$V_4(d,r)$. In eq.(\ref{V4}) it is $V_{qq}(d)$ that is subtracted, since
for $r>d $ also $V_{qq}(r)>V_{qq}(d)$ -- as will be seen in section 2. The use
of  $V_{qq}(r)$ -- corresponding to the $M_{14}M_{23}$ partition of fig.1c)
would not, therefore, be a reasonable choice for extracting a four-quark
potential from $E_4(d,r)$. It should be added that in the model, which is
developed in section 4 in order to interpret the Monte Carlo results, the
 subtraction analogous to that in eq.(\ref{V4}) is also made.
One consequence of using the quenched approximation is that the
interaction between the above meson-like quark clusters does not
contain the effect of meson-exchange between these clusters. The latter
are expected to be dominant for sufficiently large intercluster
distances. However, since the present work only discusses distances
upto seven lattice spacings (i.e. upto $\approx 0.8$fm) , the resultant
interaction should still be the dominant component.

At this point, the first objectives have now been achieved i.e. a reliable
two-quark potential $V_{qq}$ and accurate energies $V_4(d,r)$ for a series
of four-quark states. In section 4, $V_4$ is analysed in terms of a model
based on $V_{qq}$. There it is found that -- for these particular quark
configurations -- $V_4$ is {\em not} well
approximated in terms of two-body interactions between the two partitions
$A$ and $B$ in figs. 1b) and 1c). This is a situation contrary to the
corresponding many-nucleon problem. In an attempt to understand this
difference, the two-body approach is modified in two ways. Firstly, the
off-diagonal terms connecting the two configurations $A$ and $B$ are
multiplied by a factor $f$, which is dependent on the quark configuration
considered. This $f$ can be interpreted as a gluon field overlap factor.
In a second modification, the basic form of the interquark potential
$V_{qq}$ is varied.
 In section 5 convenient parametrizations of $f$ are
proposed and in section 6 some conclusions are given.

\section{The quark-quark potential}
\setcounter{equation}{0}

\vspace{1.0cm}

Even though the form of the quark-quark potential is well documented, to be
consistent in the present analysis, $V_{qq}$  will also be calculated directly
by Monte Carlo simulation. For convenience, from now on the notation
$V_{qq}=v$ will often be used. In the two-body approach of section 4, in
addition to requiring $v$ along the sides of the rectangle i.e. $v_{13},v_{24},
v_{14}$ and $v_{23}$, it will also be
necessary to know the value of $v$ between quarks at the ends of the
diagonals in fig.1 i.e. $v_{12}$ and $v_{34}$.
 These are measured directly by Monte Carlo simulation using L-shaped
 separations of the two static sources. It  is worth remembering that
 the lattice has a cubic rotational symmetry and so full rotational
 invariance is not guaranteed. It is known that at larger $r$-values
 this full rotational invariance is restored to a very good
 approximation at the present values of $\beta$.  Moreover the main features
 of the breakdown of this invariance at small $r$ can be taken into
 account by using the lattice one-gluon-exchange propagator. Thus
 a good parametrization at $\beta=2.4$ is
\be
\label{vij}
v_{ij}=-\left(\frac{e}{r_{ij}}\right)_L+b_Sr_{ij}+v_0 \ \ ;
 \ \  ( e=0.234 \ , \ b_S=0.0736\ \  {\rm and} \ \ v_0=0.542).
\ee
where it is to be understood that the $(1/r_{ij})_L$ is the lattized form
of ref.\cite{Rebbi} i.e.

\[\left(\frac{e}{r}\right)_L=\frac{e\pi}{L^{' \ 3}}\sum_{{\bf q}}
\frac{\cos(q_xr_x+q_yr_y+q_zr_z)}
{\sum_{i=1,3}\sin^2(aq_i/2)}\]
\be
\label{lattc}
aq_i=0, \ \frac{2\pi}{L'},......., \ \frac{2\pi(L'-1)}{L'} \ \
 {\rm and} \ \ {\bf q}\not =0,
\ee
where $L'=2L$, since in $SU(N)$ image sources are effectively at a distance
of $N$ times the basic lattice size $L$. Because the expression in
eq.(\ref{lattc}) only gives the energy up to an additive constant, it is
necessary -- at some particular value of ${\bf r}$ that is
large enough for lattice artefacts to be negligible -- to fix the potential to
the usual non-lattized form of eq.(\ref{vij}) containing simply $e/r_{ij}$.
 Here the point
${\bf r}=(6,0,0)$ is chosen.

               Since, for these off-axis configurations,
the difference between $v_{12}$ and $v(r_{12})$ is greatest for small $r_{ij}$,
 the off-axis configurations are calculated directly up to ($x/a=3,y/a=3$)
for $z/a=0$.
While for $x/a\ge 4$, only the on-axis potentials
$v_{MC}(x/a,0)$ are evaluated. At these large values of $r_{ij}$, the off-axis
values of $v$ are assumed to be given accurately by the above parametrization.
 The resulting two-quark potentials are shown in
table 1 as $av_L(x/a,y/a)$ and are seen to be, in general, within $1\%$ of
the exact values $v_{MC}$.
If the above string energy $b_S=0.0736/a^2$ is assumed to equal the
experimental value of 0.194 GeV$^2$ \cite{CM1,CM2},
 then the resulting lattice spacing is
$\approx 0.12$fm. This assumption that the experimental string energy can be
used in the fictional world of $SU(2)$ seems to be reasonable, since
ratios of physical quantities are very similar in $SU(2)$ and $SU(3)$
e.g. the ratio $m_{GB}/\sqrt{b_S}$, where $m_{GB}$ is the glueball mass,
is close to 3.5 in both cases \cite{chrism}.

 In section 4, it will be seen that
 it is the {\em difference} between various $v_{ij}$ that enters when
applying the two-body approach. For example, with square configurations,
it is the difference $v(x/a,x/a)-v(x/a,0)$ that is needed. Therefore, to
be consistent in section 4, $v_{MC}$ is used for $x/a<4$ and $v_L$ for
$x/a\geq 4$.

In table 1, the errors in square brackets
quoted on the $v_{MC}(x/a,y/a\not =0)$
are the ones that should be used directly with the difference with respect to
$v_{MC}(x/a,0)$ e.g. $v_{MC}(2,2)-v_{MC}(2,0)=0.1052[13]$

\vspace{1.0cm}

\section{ The four-quark potential $V_4$ from a Monte Carlo simulation}
\setcounter{equation}{0}

Since most of the techniques involved in Monte Carlo simulations are quite
standard, only the points relevant to the present calculation will be
discussed. The lattice is $16^3\times 32$ with $\beta =2.4$ as in
ref.\cite{GMP}.
 After equilibration
by a heat bath method, this lattice is then further updated by a combination
of three over-relaxation sweeps with one heat bath sweep,
 followed by a measurement of the appropriate correlation
functions. These functions are constructed from the basic lattice links
$U^0_{\mu}(n)$ that have been "fuzzed, blocked or smeared"\cite{CM1,CM2} in
the manner illustrated in fig. 2 i.e.
\[ U^0_{\mu}(n)\rightarrow U^1_{\mu}(n)=A^1_n\left[
cU^0_{\mu}(n)+\sum_{\stackrel{\pm \nu \not = \mu}{\nu\not = 4}}
 U^0_{\nu}(n)U^0_{\mu}(n+\bar{\nu})U^{0 {\dagger}}_{\nu}(n+\bar{\mu})\right],\]
\be
\label{fuzz}
 U^1_{\mu}(n)\rightarrow U^2_{\mu}(n)=A^2_n\left[
cU^1_{\mu}(n)+\sum_{\stackrel{\pm \nu \not = \mu}{\nu\not = 4}}
 U^1_{\nu}(n)U^1_{\mu}(n+\bar{\nu})U^{1 {\dagger}}_{\nu}(n+\bar{\mu}) \right]
\ \ , \ \ ....
\ee

Here the $A^i_n$ are normalisation factors chosen to project the
$U^i_{\mu}(n)$ into $SU(2)$ and $c$ is a free parameter. Recent experience
\cite{CM1,CM2} has shown
that $c=4$ is a suitable value for the present class of problems. Another
degree of freedom is the amount of fuzzing. For correlations over large
distances the greater the fuzzing the better the efficiency of the
calculation in the sense that -- the wavefunction between the quarks, generated
by connecting together  a series of fuzzed links,  has a greater overlap
with the ground state wavefunction. In some cases this has been carried to
great
lengths e.g. in ref.\cite{Booth} 110 fuzzing iterations were used, since
there the emphasis was on quark separations upto 24 lattice spacings.

In the calculation of the interquark potential, the fuzzing procedure plays
the second role of generating different paths ($P_i$) between quarks,
 for use in the variational
approach of ref.\cite{CM1,CM2}. In the latter, the correlations
\be
W^N_{ij}=<P_i|\bar{T}^N|P_j>
\ee
are calculated. Here $\bar{T}=\exp(-a\bar{H})$ is the transfer matrix for a
single time step $a$, with the Hamiltonian $\bar{H}$, the $P_{i,j}$ are
paths constructed as products of fuzzed basic links and $N$ is the number
of steps in the imaginary time direction. As shown in
ref.\cite{CM1,CM2}, a trial wavefunction $\psi=\sum_ia_i|P_i>$
leads to the eigenvalue equation
\be
\label{WN}
W^N_{ij}a^N_j=\lambda^{(N)}W^{N-1}_{ij}a^N_j.
\ee
For a single path this reduces to
\be
 \lambda^{(N)}=\frac{W^N_{11}}{W^{N-1}_{11}}=\exp(-aV_0),
\ee
where $V_0$ is the potential of the quark system being studied. Unfortunately,
in this single path case, $N$ needs to be large and this can lead to
unacceptable error bars on
the value of $V_0$ extracted. However, if a few paths are taken, it is found
that $N$ need only be small $(N<5)$ to get a good convergence to $V_0$ with
small error bars. A further advantage of this approach is that not only can
the lowest eigenvalue be extracted but also higher ones.

During the Monte Carlo simulation, the correlations $W_{ij}(r)$ appropriate
for extracting the two-quark potential $V_{qq}(r)$ and the correlations
$W_{ij}(d,r)$ for
$V_4(d,r)$ are evaluated at the same time. For $W_{ij}(r)$ three paths are
generated by the three different fuzzing levels 12,16 and 20. On the other
hand, for the $W_{ij}(d,r)$ only the level 20 is kept, with the variational
basis being the two configurations $A$ and $B$ in figs. 1b) and 1c). These
lead to the Wilson loops in fig.3.

 In the case of squares (i.e. $r=d$), eqs.(\ref{V4},\ref{WN}) give the
potential energy of the ground state as
\be
\label{le0}
E_0=aV_4(d,d)=\log\left[\frac{W_{11}^{N-1}(d,d)+W_{12}^{N-1}(d,d)}
{W_{11}^{N}(d,d)+W_{12}^{N}(d,d)}\right] -2E(d)
\ee
and that of the first excited state
\be
\label{le1}
E_1=\log\left[\frac{W_{11}^{N-1}(d,d)-W_{12}^{N-1}(d,d)}
{W_{11}^{N}(d,d)-W_{12}^{N}(d,d)}\right] -2E(d),
\ee
where $E(d)$ is the result of diagonalising the 3*3 variational basis
for $V_{qq}$. In these equations, $N$ should tend to $\infty$. However, in
practice it is found to be sufficient to have $N<5$ for extracting accurate
values for $E_0$ and $E_1$.

The results are shown in fig.4 and in the first three columns of table 2.
For the squares, these are the outcome of 6400 sweeps
 divided into 20 blocks  of 320 sweeps. The sweeps were carried out in
groups of four -- three over-relaxation sweeps  plus one heat-bath update
sweep -- followed by a correlation measurement. The measurements ( 1600 in all)
were kept in blocks in order to estimate the errors associated with each value
of $E_0$ and $E_1$. For rectangles, 20 blocks of only 160 sweeps were
 analysed. Since, for each block of measurements, $E_0$ and $E_1$ are
extracted directly  as the {\em difference} between two correlated
quantities, their overall errors are expected to be smaller
than those for the $E_4$ and $v(d)$ in eq.(\ref{V4}) separately.
In practice, it was found that in eq.(\ref{WN}) $N=2$ or 3 were adequate
to extract  results that were consistent within the errors.
 Initial test runs were carried out on the Cray X-MP at the
Rutherford-Appleton Laboratory (UK) with the subsequent development and
production
runs being performed on the corresponding Helsinki machine. In the case of
squares, each of the 20 blocks took about 8 hours of Cray CPU time.
For comparison, fig.4
also shows the corresponding 1985 results of ref.\cite{Ohta} as the two points
at $r/a\approx 1.25$ and 2.5 -- remembering that the lattice spacing for
ref.\cite{Ohta} is $a'\approx 0.15$fm $\approx 1.25a$.
It is seen that the present calculation gives
energies that have error bars which only become significant for
$r/a\geq 7$, whereas the 1985 work was unable to generate any meaningful
numbers beyond $r/a'=2$. In fact, for the present purpose, neither
of the 1985 points is of use, since the $r/a'=1$ result could
well suffer from lattice artefacts -- a problem afflicting all calculations of
configurations involving a single lattice spacing -- and the $r/a'=2$
result has error bars that are too large for the analysis to be carried out
in the next section. However, the following should be added in defence of
ref.\cite{Ohta}. Firstly, their calculation was for $SU(3)$ and so it was at
least
an order of magnitude more demanding on CPU time. Secondly, it should be
remembered that the energies of interest ($E_{0,1}$) are very small compared
with the total four-quark energy $E_4$ in eq.(\ref{V4}). For example, at
$r=d=4a$ the value of $aE_0=-0.050(1)$ is obtained from the difference
between $aE_4$ and $2av(d)$, which is 1.505--1.555 i.e. the error quoted
on $aE_0$ is less than $0.1\%$ of $aE_4$.

One of the most outstanding features of the results in table 2 is that for
\underline{$r=d$} the value of $aE_0$ decreases smoothly in magnitude
from --0.07
for $r=d=1$ to --0.04 for $r=d=6$. However, for the few cases where $r\not=d$
the value of $E_0$ is {\em an order of magnitude smaller} than
the adjacent $r=d$ cases e.g. $E_0(2,3)\approx -0.006$, whereas $E_0(2,2)$
and $E_0(3,3)$ are $\approx -0.055$. This result is reminiscent of the
conclusion found with the flux-tube model \cite{Paton} and the
ansatz made in the flip-flop model of refs.\cite{lenz}.  In both of these
models the interaction between the two separate two-quark partitions of
fig.1 is very small ( in fact zero in the flip-flop model) except when
the unperturbed energies of the two partitions is the same i.e. at
$r=d$ in the case of rectangles.

It should be added that the potentials for configurations (1,1) and (7,7)
may be misleading, since the former could suffer from lattice artefacts and
the latter from finite size lattice effects. Here they are included for
completeness in anticipation\cite{GMPS} that eventually these two points will
be checked by varying $\beta$ and increasing the lattice size from $L=16$ to
something larger.

For completeness, it should be pointed out that the values of
$v_{MC}(x/a,y/a\not =0)$ in table 1 are also calculated with a single fuzzing
of 16 with $c=4.0$.

Throughout this work it was checked that there was no statistically
significant correlation between blocks of measurements.

\vspace{0.5cm}

\section{ The two-body approaches}
\setcounter{equation}{0}

In the previous two sections accurate values of the interquark potential
$v_{ij}$ and the four-quark ground state binding energy $E_0$ and excited state
energy $E_1$ have been extracted using Monte Carlo simulation. In this
section, attempts will be made to understand the $E_{0,1}$ in terms of the
$v_{ij}$. With this aim in mind several models will be suggested, in which
 the basic two-quark potential gets increasingly modified. The first model uses
simply
the original form of $v_{ij}$ in eq.(\ref{vij}), whereas the second model
modifies this form into another purely two-body interaction. The third model
introduces implicitly
multi-quark effects generated by the gluon-field that are in addition to those
 already contained in the $v_{ij}$.

\vspace{0.5cm}

a) Unmodified two-body approach.

\vspace{0.5cm}

Once the quark-quark potential $v_{ij}$ is known and, if it is assumed to be
the
{\em only} interaction between the quarks, then the energy of a multiquark
system can be readily calculated -- provided the wavefunction for that system
is expressed in terms of a sufficient number of basis states. For the present
situation,
 the most obvious choices for such states are $A$ and $B$ in figs. 1 b) and c).
In this extreme two-body approach, since the presence of the gluon fields
 have been explicitly removed -- their only effect now being in the colour
indices of the quarks --, the states $A$ and $B$ form a complete but
non-orthogonal basis. This implicitly assumes in fig.1 the quark$(q)$,
antiquark$(\bar{q})$ assignment $q(1)\bar{q}(3)q(2)\bar{q}(4)$. In $SU(2)$
the other quark assignment $q(1)q(3)\bar{q}(2)\bar{q}(4)$ is numerically
equivalent and so leads to nothing new.
In this approach, states are excluded in which the
gluon fields are excited.  As discussed earlier in refs.\cite{GMP,MGP},
the energies ($E'_i$) of this static four-quark system can be extracted from
the eigenvalues of the Hamiltonian
\be
\label{Ham}
\left({\bf V}-\lambda_i {\bf N}\right)\Psi_i=0
\ee
where
\be
\label{NV}
{\bf N}=\left(\begin{array}{ll}
1&1/2\\
1/2&1\end{array}\right)\ {\rm and}\ {\bf V}=\left(\begin{array}{cc}
v_{13}+v_{24} & V_{AB}\\
V_{BA}&v_{14}+v_{23}\end{array}\right).
\ee
Several points need explaining in these equations.

1) The off-diagonal matrix element $N_{12}=<A|B>=1/2$ shows the
non-orthogonality of the $A,B$ basis. At this stage, the lack of orthogonality
could have been avoided by using the basis $A\pm B$. However,
later it will be seen that the $A,B$ basis is in fact more convenient and
suggestive, when the gluon fields are reintroduced in a more explicit manner.

2) In section 2, the interquark potential $V_{qq}(ij)=v_{ij}$ was extracted
 as the potential energy of two quarks. In order
to evaluate potential energy matrix elements, a further assumption is needed
 concerning the colour structure  of $v_{ij}$. Here the usual
identification
\be
\label{vcol}
V_{ij}=-\frac{1}{3} {\bf \tau}_i.{\bf \tau}_j v_{ij}
\ee
will be made, where the ${\bf \tau}_i$ are the Pauli spin matrices
appropriate for $SU(2)$. This choice ensures, for a meson-like state
$[ij]^0$, that
\be
<[ij]^0|V_{ij}|[ij]^0>=v_{ij}.
\ee
Strictly speaking, the form in eq.(\ref{vcol}) is only true in the weak
coupling limit of one-gluon-exchange, since this has replaced the local
gauge invariance -- ensured by the series of $U$-links of
eq.(\ref{fuzz}) connecting the two quarks -- by the global gauge
invariance reflected by the ${\bf \tau}_i.{\bf \tau}_j$ factor.

3) In the spirit of the {\em global} colour invariant two-body
potential picture expressed by eq.(\ref{vcol}),  the assumption leading
to this equation  may be relaxed to
include an arbitrary potential $w_{ij}$ with the structure of a colour singlet
exchange i.e.
\be
\label{modv}
V_{ij}=-\frac{1}{3} {\bf \tau}_i.{\bf \tau}_j(v_{ij}-w_{ij}) + w_{ij}.
\ee
Then additional contributions will appear in all matrix elements
of ${\bf V}$. This possibility will be discussed later.

4) With the choice of $V_{ij}$ in eq.(\ref{vcol}), the off-diagonal
potential matrix element becomes
\be
\label{AVB}
<A|V|B>=V_{AB}=V_{BA}=
\frac{1}{2}\left(v_{13} +v_{24} +v_{14}+v_{23} - v_{12}-v_{34} \right).
\ee
Since the following discussion only involves quark configurations in the
rectangular geometries of fig.1, it is convenient to use the notation
\be
\label{not}
v_{13}=v_{24}=v_{s_1}, \ \ v_{14}=v_{23}=v_{s_2} \ \ {\rm and} \ \
v_{12}=v_{34}=v_d,
\ee
where the indices $s_i$ and $d$ refer to the sides and the diagonal of the
rectangles.

Even though the form of eq.(\ref{AVB}) is derived in the one-gluon-exchange
limit, it is now assumed that a more realistic model emerges if the
$v_{ij}$ are taken to be the complete potential of eq.(\ref{vij}) and
not just the one-gluon-exchange component. This clearly has the correct
form when the distance between the two two-quark clusters of fig.1 are far
apart, since in this case the only interactions are those {\em within}
 the separate clusters.

The $2\times 2$ Hamiltonian of eq.(\ref{Ham})
is easily diagonalised to give the eigenvalues
$\lambda _{0,1}$. Since it is the binding energy $E'_i$ of the four-quark
system that is of interest, and also that was extracted in section 2 from the
Monte Carlo simulation, the internal energy of the meson-like state
with the lowest energy (i.e. $2v_{s_1}(d)$ with $d\leq r$) is now subtracted
from the $\lambda_i$ to give
\be
\label{E'}
 E'_i=\lambda_i-2v_{s_1}(d)
\ee
in analogy with eq.(\ref{V4}).

Therefore, in this simpliest version of the two-body approach, the $E'_i$'s
should correspond to the $E_i$'s from the Monte Carlo simulation -- a
comparison which is made in table 2 and fig.4. In this table, the significance
of the "$f=1$" label on the $E'_i$ will be apparent later. Since the values
of $E_0$ for squares are much larger than those for neighbouring rectangles,
only the results for squares are shown in fig.4. In these cases
\be E'_0(f=1)=-\frac{2}{3}(v_d-v_{s_1}) \ \ {\rm and} \ \
 E'_1(f=1)=2(v_d-v_{s_1})=-3E'_0,
\ee
where from table 1, $v_d=v(r/a,r/a)$ and $v_{s_1}=v(r/a,0)$. In addition,
the corresponding normalised wavefunctions are
$\psi(E'_0)=\frac{1}{\sqrt{3}}|A+B>$ and $\psi(E'_1)=|A-B>$.

Two comments should be made on these results.

1) For the smallest rectangles the agreement between $E_i$ and $E'_i$ is best.
This is reasonable, since it is expected that at small interquark distances
perturbation theory is adequate i.e. the lowest order gluonic effects are
already incorporated correctly into the interquark potential.

2) As the rectangles get larger the differences between the $E_i$ and
the $E'_i$ grow until $E'_0$ is more than three times $E_0$, and $E'_1$
more than seven times $E_1$, for the largest squares
$r\approx 7a\approx 0.8$fm. Since $E'_0$ is too attractive
and $E'_1$ too repulsive, this suggests that the off-diagonal matrix
element $V_{AB}$ is too large.

The main conclusion to be drawn from the above is that the two-body
potential of eq.(\ref{vcol}) does {\em not} give the potential energy of the
four-quark system -- the indication being that the off-diagonal
potential energy $V_{AB}$ is too large.

\vspace{1.0cm}

b) The effect of a modified two-body approach as in eq.(\ref{modv}).

\vspace{0.5cm}

In the previous subsection, it was assumed that the potential $v_{ij}$
extracted from the Monte Carlo simulation has the colour structure of
eq.(\ref{vcol}). However, as mentioned above, a more general assumption
-- based on an unjustified {\em global} gauge invariance -- is
that of eq.(\ref{modv}), in which an arbitary potential $w_{ij}$,
with the structure of a colour singlet exchange,
has been introduced.  Using the notation of eqs.(\ref{modv},\ref{not}),
it can be shown that for squares

\[E'_0(w_{ij})=-\frac{2}{3}[v_d-v_{s}-(w_d-w_{s})]+2(w_s+w_d)=
E'_0(f=1)+\frac{4}{3}(2w_d+w_s)\]
\be
\label{e'1}
E'_1(w_{ij})=2[v_d-v_{s}-(w_d-w_{s})]+2(w_s+w_d)=
E'_1(f=1)+4w_s
\ee

Therefore, this possibility alone cannot give  a direct understanding of both
$E_0$ and $E_1$, since the effect of $w_{ij}$ acts in the same way for both
$E'_0$ and $E'_1$, whereas $E'_0$ needs a repulsive mechanism and $E'_1$
an attractive one.

\vspace{1.0cm}

c) The two-body potential modified by multi-quark effects.

\vspace{0.5cm}

In the two previous models it is assumed that all of the gluonic effects
are incorporated into the two-body potential $v_{ij}$. However, this is
clearly an oversimplification that is at best only applicable in situations
where perturbation theory holds -- namely at short distances -- as already
noted in the discussion of fig.4. In more realistic models, the QCD coupling
is sufficiently strong to constrain the gluon field into flux-tubes connecting
the quarks in a given meson -- as visualised by the wavy lines between
the quarks in states $A$ and $B$ in fig.1. Therefore, the overlap of states
$A$ and $B$, i.e. $N_{12}=<A|B>$, is not simply the colour recoupling factor
of 1/2, but should also involve the lack of overlap of the gluon fields in
states $A$ and $B$. This can be incorporated by introducing an entity $f$,
which simply multiplies the original $N_{12}$, and which is an unknown
function of the position coordinates of the four quarks. With this
interpretation of $f$ as a gluon field overlap factor, it is also necessary
to multiply the off-diagonal potential matrix element $V_{AB}$ of
eq.(\ref{AVB}) by the {\em same} factor $f$. This factor must be the same
in $N_{12}=f/2$ and $V_{12}=fV_{AB}$ otherwise the binding energies
$E'_{0,1}$, extracted from eq.(\ref{E'}) after diagonalising
\be
\label{Hamf}
\left[{\bf V}(f)-\lambda_i(f) {\bf N}(f)\right]\Psi_i=0,
\ee
with
\be
\label{NVf}
{\bf N}(f)=\left(\begin{array}{ll}
1&f/2\\
f/2&1\end{array}\right)\ \ {\rm and}\ \ {\bf V}(f)=\left(\begin{array}{cc}
v_{13}+v_{24} & fV_{AB}\\
fV_{BA}&v_{14}+v_{23}\end{array}\right),
\ee
would be dependent on the self-energy term $v_0$ in the form of $v_{ij}$
given in eq.(\ref{vij}) -- which would be unphysical.

The two equations (\ref{Hamf} and \ref{NVf}) are the basis of the following
analysis. They give a procedure for extending the model of eqs.(\ref{Ham},
\ref{NV}), which was justified in the weak coupling limit, into the
domain beyond one-gluon-exchange. The off-diagonal potential matrix
element performs this extension in two ways. Firstly, even though $V_{AB}$
in eq.(\ref{AVB})
still has the same algebraic structure in terms of the $v_{ij}$ as dictacted
by the one-gluon-exchange limit, the $v_{ij}$'s themselves are the full
two-quark potential of eq.(\ref{vij}). Secondly, in the off-diagonal
correlation $W_{12}$ of fig.3, the one-gluon-exchange
model suggests the presence of the overall multiplicative factor $f$ due
to gluon exchange within the {\em initial} and {\em final} states $A$ and $B$
at euclidean times $T=0$ and $\infty$. In this interpretation, the terms in
$V_{AB}$ arise during the period of propagation between $T=0$ and $\infty$.

The strategy is now to adjust $f$ to get an exact fit to $E_0$ or $E_1$. In
the case of squares, since
\be
\label{e01}
E'_0=\frac{-f}{1+f/2}(v_d-v_s) \ \ {\rm and} \ \
E'_1=\frac{f}{1-f/2}(v_d-v_s),
\ee
the appropriate expressions are
\be
\label{f's}
f(E_0)=\frac{E_0}{v_s-v_d-E_0/2} \ \ {\rm and}    \ \
f(E_1)=\frac{E_1}{-v_s+v_d+E_1/2}
\ee
for fitting $E_0$ and $E_1$ respectively. In table 3 are given the values
of $E_{0,1}$ and $f(E_{0,1})$ from these equations. For the non-square
rectangles the corresponding equations are somewhat more complicated.

The first point to note from these results is that all values of $f$
are less than unity as is expected from the interpretation of $f$ as
a gluon field (lack-of-) overlap factor. In addition, this idea is
supported by the fact that the values of $f$ decrease as the quarks
get further apart.

It is also seen that $f(E_0)$ and $f(E_1)$ are quite close to each other --
suggesting that a single compromise value of $f$ could give a reasonable
description of both $E_0$ and $E_1$. This is demonstrated in the next
subsection. The existence of a compromise value of $f$
 is even more apparent in fig.5
and also table 3, where the value of $f(E_0)$ is used to give $E'_1$ i.e.
$E'_1[f(E_0)]$ and $f(E_1)$ to give $E'_0$ i.e. $E'_0[f(E_1)]$.
In this notation $E'_0[f(E_0)]=E_0$ and $E'_1[f(E_1)]=E_1$.
There it is seen that, to within $15\%$, $E'_1[f(E_0)]\approx E_1$ and
$E'_0[f(E_1)]\approx E_0$. The actual values of $f(E_{0,1})$ are shown
in fig.6.

\vspace{1cm}

e) A compromise value for the gluon overlap factor $f$.

\vspace{1.0cm}

The previous subsection suggests that for each geometry a {\em single}
value of $f$ could give a reasonable description of both $E_0$ and $E_1$.
Here one such possibility is given by finding that value of $\bar{f}$ which
minimizes the expression
\be
\label{dmin}
D(\bar{f})=\left(\frac{E_0-E'_0(\bar{f})}{\Delta E_0}\right)^2+
       \left(\frac{E_1-E'_1(\bar{f})}{\Delta E_1}\right)^2,
\ee
where the $\Delta E_i$ are the errors quoted for the $E_i$ in table 2.
The results are shown in table 4.

It is seen that indeed a single value of $f=\bar{f}$ suffices to explain
reasonably well
both energies. This is a non-trivial observation, since it indicates that
the parametrization suggested in eqs.(\ref{Hamf},\ref{NVf}) contains
the most important features of the more precise lattice calculation.

It should be added that the extraction of a compromise value of $f$ is
not simply a curiosity, since any model that needs different values of
$f$ for $E_0$ and $E_1$ would be more difficult to use in practice for
more complicated multi-quark systems.

At this point, even though for each geometry a single value of $f=\bar{f}$
gives  values
of $E'_i$ that are in reasonable agreement with the $E_i$, it might be asked
about the remaining small differences. Two possibilities are now open.
Firstly, the lattice energies $E_i$ may not be sufficiently accurate due to
finite lattice size and scaling uncertainties -- a point now being
studied in ref.\cite{GMPS}.
Secondly, the  parametrization in eqs.(\ref{Hamf},\ref{NVf}) may
be inadequate. One possibility would be to combine the notion of a gluon
overlap factor $f$ with a generalized form for the two-quark potential, since
this could introduce more free parameters. However, in this direction the only
 suggestion made in this paper is the form given in eq.(\ref{modv}), for which
there is no basic justification when combined with an overlap factor $f$.
Perhaps a more fruitful approach, more in line with the step from
eqs.(\ref{Ham},\ref{NV}) to eqs.(\ref{Hamf},{\ref{NVf}), would be to perform
a {\em two-gluon} exchange calculation to see what new terms arise and to
then be guided by this in making an improved parametrization. Another
possibility is that any model based only on states $A$ and $B$ in fig.1
is incomplete and that other states are necessary in addition.

\vspace{1.0cm}

In this section various models have been proposed in an attempt to
understand the results of the Monte Carlo simulation. The main outcome --
summarised
in fig.6 -- is the emergence of a function $f$ that depends on the
coordinates of the four
quarks involved. This shows that the model based on purely two-quark
interactions needs to be modified considerably -- essentially by the factor
$f$. This observation is in itself
of much interest, but at this stage it is not clear
how the effect can be incorporated into more realistic situations in which
the quarks are not so restricted in their geometry. It is the purpose
of the next section to tackle this problem by first studying how $f$
can be parametrized.

\vspace{1.0cm}

\section{ Parametrizations of the gluon-field overlap factor $f$ }
\setcounter{equation}{0}

In the previous section, models were introduced in an attempt to understand
the ground state binding energy $(E_0)$ and excited state energy $(E_1)$
emerging from a Monte Carlo simulation, in which four quarks were at the
corners
of a rectangle. These models are summarised by
eqs.(\ref{Hamf},\ref{NVf}).
In both cases, for each quark configuration, the energies $E_{0,1}$ are
described in terms of a function $f$ of the four quark positions. As it stands,
this is not particularly useful unless $f$ can be parametrized in some
sensible and convenient manner. In the literature, two such parametrizations
have been suggested. In refs.\cite{Morimatsu,Morimatsu2}
-- motivated by strong coupling arguments -- the phenomenological
form is taken to be
\be
\label{f1}
f_1=\exp[-\alpha b_S S],
\ee
where $b_S=0.0736$ is the string tension dictated by the interquark
potential of eq.(\ref{vij}) and $S$ is the minimal area of the surface
bounded by the straight lines connecting the quarks and antiquarks.
The other form -- the one proposed in \cite{Masud} -- is
\be
\label{f2}
f_2=\exp[-\frac{kb_S}{6}\sum\limits_{i<j}r^2_{ij}]
\ee
i.e. the cut-down is governed by the average of the six links
present in a $q^2\bar{q}^2$ system.
In eqs.(\ref{f1},\ref{f2}) the $\alpha$ and $k$ are at present free
parameters to be determined later.
Both of these parametrizations of $f$ accommodate the two
extreme models of:

1) Weak coupling, which has $f=1$ when all $r_{ij}=0$ .

2) Strong coupling, which has $f=0$ when any $r_{ij}\rightarrow \infty$.

With squares $(r=d)$, for which the most accurate values of $f$ exist,
$k$ equals $3\alpha/4$. One measure of how meaningful these parametrizations
really are, is given by extracting $\alpha $ and $k$ for each quark
configuration. The hope would then be that, for squares, $\alpha$ (and
therefore
$k$) would be independent of the separate configurations. Only the few points
for non-square rectangles $(r\not = d)$ would be able to distinguish between
$f_1$ and $f_2$. In table 5 and fig. 7, $k(E_0),k(E_1)$ and $\bar{k}$ ( and the
corresponding $\alpha$'s) are given for the $f(E_0),f(E_1)$ and $\bar{f}$
from tables 3 and 4.

As seen in fig. 7, $k(E_0)$ for the squares appears to decrease slowly from
about
0.7 to about 0.5 as the sizes of the squares increase from (2*2) to (7*7),
whereas for non-squares $k(E_0)$ appears to be stable at about 0.7$\pm$0.1.
On the other hand, $k(E_1)$ and $\bar{k}$ decrease somewhat less and also the
square and non-square values are consistent with each other. As mentioned
earlier, the points involving a single lattice spacing [i.e. (1,2)] could
suffer from
 lattice  artifacts and those for the largest number of spacings [i.e. (7,7)]
could suffer from finite size effects. The indications from this discussion
are that $\bar{k}$ is a suitable compromise value that results in
a reasonable fit to both $E_0$ and $E_1$.

In fig. 8 an attempt is made to see whether there is a
difference between the parametrizations $f_{1,2}$ of $f$ in
eqs.(\ref{f1},\ref{f2}). However, in the two cases discussed, it is
found that $k_0$ is not significantly different to $3\alpha_0/4$, and
like-wise with  $\bar{k}$ and $3\bar{\alpha}/4$ .

\vspace{0.5cm}

\section{ Conclusions }
\setcounter{equation}{0}
 In this paper it has been shown that -- in $SU(2)$ -- the binding energy
 of four quarks situated at the corners of a series of rectangles can be
 accurately extracted from a Monte Carlo simulation using a lattice
$16^3\times 24$ with $\beta=2.4$. Squares with sides upto six lattice spacings
( i.e. $\approx 0.7$fm) could  be treated reliably  -- with larger squares
possibly suffering from finite lattice size effects. The resultant binding
energies for these squares were, in all cases, between --0.04 and --0.007
in lattice spacing
units i.e. between --70 and --110MeV. In addition, the corresponding first
excited states
were between 80 and 300MeV -- see table 2.
For non-squares, the ground state binding energy was at least an order of
magnitude
smaller, showing that the interaction is strongly peaked at $r\approx d$.

In an attempt to understand these four quark binding energies a model was
constructed that expressed these energies in terms of the conventional two-body
interquark potential. However, this procedure, even though it generated
energies
 that were in
line with the Monte Carlo energies for rectangles with the smallest number
of lattice spacings, gave far too much binding for the larger
rectangles -- see table 2 and figure 4. In other words a simple two-body
potential picture is {\em not} adequate for explaining the Monte Carlo data.
 To reduce this over-binding,
a factor $f$ was introduced into the model. This factor was interpreted
as a gluon field overlap factor  and was dependent only on the relative
positions of the four quarks. In this way -- with a single value
of $f$ for each quark geometry -- both the ground state and first excited state
energies could be fitted reasonably well -- see table 4 and figure 5.

To make  use of the factor $f$ in dynamical calculations,
it would be necessary that it be conveniently
parametrized. Here two forms were suggested -- see
eqs.(\ref{f1},\ref{f2}) -- that are essentially gaussian in character and
are, in fact, equivalent for squares. In this way a reasonable description
of the $f$ results -- with the non-squares being unable to distinguish
between the two suggested forms for $f$ -- see figure 8.

 Of course, the next question to be addressed concerns the universality
of the parametrizations of $f$. This is now under study \cite{GMPS}
and involves the determination of four quark binding energies in
non-rectangular geometries.

The authors wish to acknowledge that these  calculations were performed on
the CRAY X-MP machines at both Helsinki and RAL (UK) - computer time on
the latter being financed by a grant from SERC. One of the authors (A.M.G.)
also wishes to thank M.E.Sainio for many useful discussions.

\newpage

\vspace{1.0cm}

Table 1. The interquark potential $V_{qq}$, as a function of $(x,y)$, for a
series of approximations -- a preliminary version of this table can be found
in ref.\cite{Hunt}.

a) $v_{MC}(x/a,y/a)$ corresponds to the present Monte Carlo simulation.

b) $v_L(x/a,y/a)$ is from eq.(\ref{lattc}) -- adjusted to fit the asymptotic
value of 0.9446 at $(x,y)=(6,0)$ -- obtained from eq.(\ref{vij}) with
$\left(\frac{e}{r_{ij}}\right)_L$ replaced by
$\left(\frac{e}{r_{ij}}\right)$.

\[ \begin{array}{|c|c|c|} \hline

(x/a,y/a)& av_{MC}(x/a,y/a)& av_L(x/a,y/a)  \\ \hline
(1,0) & 0.3736(1) & 0.3634 \\
(1,1) &0.4886[4]  & 0.4846 \\
(2,0) & 0.5627(3) &0.5639  \\
(2,1) & 0.6016[4]& 0.6017  \\
(2,2) & 0.6679[13]&  0.6684  \\
(3,0) & 0.6815(5) & 0.6825 \\
(3,1) & 0.6998[12]  &0.7004  \\
(3,2) &  0.7432[12] &0.7431  \\
(3,3) &  0.7977[24] &0.7998  \\
(4,0) & 0.7752(8) &0.7774  \\
(4,4) &         & 0.9174  \\
(5,0) & 0.8596(13) &0.8631  \\
(5,5) &         & 1.0295 \\
(6,0) & 0.9391(17)&\underline{0.9446}  \\
(6,6) &         & 1.1388 \\
(7,0) &  1.016(3)  &1.0237  \\
(7,7) &          &1.2464  \\ \hline
\end{array}  \]

\vspace{1.0cm}

\newpage

\vspace{1cm}

Table 2. The two lowest eigenvalues for squares and rectangles
as a function of $(d/a,r/a)$ -- in units of the lattice spacing
$a=0.12$fm.

a) $E_0$ and $E_1$ are from the Monte Carlo simulation.

b) As will be discussed in the next section $E'_0(f=1)$ and $E'_1(f=1)$
are the

two lowest eigenvalues from a model based solely on two-quark
interactions.

\begin{center}
\vspace{1cm}

\begin{tabular}{|c|c|c|c|c|}  \hline
$(\frac{\displaystyle d}{\displaystyle a},\frac{\displaystyle r}
{\displaystyle a})$
& $E_0$ & $E_1$ & $E'_0(f=1)$ & $E'_1(f=1)$ \\ \hline
(1,1) &--0.0695(5) &0.184(1) &--0.0767 &0.230   \\
(1,2) &--0.0026(2) &0.505(1)&--0.0036&0.560   \\
(1,3) &--0.0003(3) &0.765(3) &--0.0005 &0.846    \\
(1,4) &0.0003(2) &0.98(1) &--0.0002 &1.092   \\
(2,2) &--0.0582(2) &0.142(1) &--0.0701 &0.210 \\
(2,3) &--0.0056(3) &0.324(3) &--0.0123&0.411 \\
(3,3) &--0.054(1) &0.117(1) &--0.0775 &0.232 \\
(3,4) &--0.006(1) &0.255(5) &--0.025&0.395 \\
(4,4) & --0.050(1) &0.096(1) &--0.093 &0.279 \\
(5,5) & --0.047(1) &0.072(3)&--0.111 &0.333 \\
(6,6) & --0.038(3) &0.055(1) &--0.129 &0.387 \\
(7,7) & --0.024(3) &0.041(3)&--0.149&0.446\\
\hline
\end{tabular}
\end{center}
\vskip 1.0cm

\newpage

\vspace{0.5cm}

Table 3.  The $f(E_{0,1})$ from eqs.(\ref{f's}) and
$E'_1[f(E_0)]$, $E'_0[f(E_1)]$.

\vspace{0.5cm}

\begin{tabular}{|c|c|c|c|c|c|c|}  \hline
$(\frac{\displaystyle d}{\displaystyle a},\frac{\displaystyle r}
{\displaystyle a})$
& $E_0$ & $E_1$ & f($E_0$)   & f($E_1$) &$E'_1[f(E_0)]$&$E'_0[f(E_1)]$ \\
\hline
(1,1) &--0.0695(5) &0.184(1) &0.866(9)  & 0.889(3)&0.176(3)&--0.0708(2)\\
(1,2) &--0.0026(2) &0.505(1)& 0.84(3) & 0.869(3)&0.495(10)&--0.00279(2)\\
(2,2) &--0.0582(2) &0.142(1) &0.765(3) &0.806(3)&0.130(1)&--0.0604(1)\\
(2,3)& --0.0056(3) &0.324(3) &0.63(2) & 0.75(1)&0.297(4)&--0.0077(2)\\
(3,3) &--0.054(1) &0.117(1) &0.61(1)  &0.670(4)&0.101(3)&--0.058(1) \\
(3,4) &--0.006(1) & 0.255(5) &0.42(4) & 0.59(2) &0.222(6)&--0.011(1)\\
(4,4) & --0.050(1) &0.096(1) &0.43(1)  &0.510(4)&0.078(2)&--0.057(1)  \\
(5,5) & --0.047(1) &0.072(3)&0.33(1) &0.36(1)&0.066(2)&--0.050(1)\\
(6,6) & --0.038(3) &0.055(1) &0.22(2) &0.248(4)&0.047(5)&--0.043(1)\\
(7,7) & --0.024(3) &0.041(3)&0.11(1) &0.17(1) &0.027(4)&--0.035(3)\\
\hline
\end{tabular}

\vspace{0.5cm}

\newpage

\vspace{1.0cm}

Table 4. The two lowest eigenvalues as a function of $(d/a,r/a)$.

a) $E_0$ and $E_1$ are from the Monte Carlo simulation given in table 2.

b)  $E'_0(\bar{f})$ and $E'_1(\bar{f})$ -- the model energies using the
compromise value of $\bar{f}$.

c) The compromise value of $\bar{f}$.

\begin{center}
\vspace{1cm}

\begin{tabular}{|c|c|c|c|c|c|} \hline
$(\frac{\displaystyle d}{\displaystyle a},\frac{\displaystyle r}
{\displaystyle a})$
& $E_0$ & $E_1$ & $E'_0(\bar{f})$ & $E'_1(\bar{f})$& $\bar{f}$ \\ \hline
(1,1) &--0.0695(5) &0.184(1) &--0.0707(2) &0.183(1)& 0.887(3)   \\
(1,2) &--0.0026(2) &0.505(1)&--0.0028(1)&0.505(1)&0.868(3)   \\
(2,2) &--0.0582(2) &0.142(1) &--0.0594(2) &0.136(1)&0.787(4) \\
(2,3) &--0.0056(3) &0.324(3) &--0.0070(2)&0.314(4)&0.709(15) \\
(3,3) &--0.054(1) &0.117(1) &--0.058(1) &0.116(1)&0.666(4) \\
(3,4) &--0.006(1) &0.255(5) &--0.009(1)&0.242(6)&0.53(3) \\
(4,4) & --0.050(1) &0.096(1) &--0.056(1) &0.094(1)&0.502(4) \\
(5,5) & --0.047(1) &0.072(3)&--0.048(1) &0.067(2)&0.34(1) \\
(6,6) & --0.038(3) &0.055(1) &--0.043(1) &0.055(1)&0.247(5) \\
(7,7) & --0.024(3) &0.041(3)&--0.031(2)&0.036(4)&0.15(1)\\
\hline
\end{tabular}
\end{center}
\vskip 1.0cm

\newpage

\vspace{0.5cm}

Table 5. The values of  $k(E_0),k(E_1)$ and $\bar{k}$ corresponding  to the
$f(E_0),f(E_1)$ and $\bar{f}$ from tables 3 and 4. The numbers [...] are the
values of $3\alpha/4$ for the non-square rectangles.

\vspace{0.5cm}
\begin{center}
\begin{tabular}{|c|c|c|c|}  \hline

$(\frac{\displaystyle d}{\displaystyle a},\frac{\displaystyle r}
{\displaystyle a})$ & $k(E_0)$ &$k(E_1)$ &$\bar{k}$
\\ \hline
(1,1)& 1.47(10)&1.20(3)&1.22(4)\\
(1,2)& 0.71(15) & 0.57(2)&0.58(1)  \\
&[0.9(2)]&[0.72(2)]&[0.72(2)]\\
(2,2)&0.68(1)& 0.55(1)&0.61(1) \\
(2,3)& 0.72(5) &  0.45(2)&0.54(3) \\
& [0.78(5)]&[0.49(2)]& [0.58(4)]  \\
(3,3)& 0.57(3)& 0.45(1)&0.46(1) \\
(3,4)& 0.71(8)&  0.43(3)& 0.51(4) \\
&[0.74(8)]& [0.45(3)]&[0.54(5)] \\
(4,4)& 0.53(1)& 0.43(1)& 0.44(1)  \\
(5,5)& 0.45(1)& 0.42(1)&0.44(1)  \\
(6,6)& 0.43(3)& 0.40(1)&0.40(1) \\
(7,7)& 0.45(3)& 0.37(1)&0.40(2) \\
\hline
\end{tabular}

\end{center}

\newpage

\hspace{0.1cm} {\bf Figure Captions}
\vspace{1cm}

Fig.1  a) The basic four-quark configuration.

b) The meson-like partition   $A=M_{13}M_{24}=[q_1q_3]^0[q_2q_4]^0$.

c) The meson-like partition $B=M_{14}M_{23}=[q_1q_4]^0[q_2q_3]^0$.

\vspace{1cm}

Fig.2 a) The basic link (solid line) between two neighbouring spatial
points (12) is augmented by a combination of basic links involving other
neighbouring spatial lattice points (34) -- to give a singly fuzzed
link (dashed line).

b) The singly fuzzed link is replaced by a doubly fuzzed link (wavy line)
.... and so on.

\vspace{1cm}

Fig.3 The correlations $W_{11}$ and $W_{12}$.

\vspace{1cm}

Fig.4  $E_{0,1}$ from Monte Carlo simulation and $E'_{0,1}(f=1)$ from
the 2*2 model to be discussed in section 4.
Also the results of ref.\cite{Ohta} are shown.

\vspace{1cm}

Fig.5. The model energies $E'_1[f(E_0)]$ and $E'_0[f(E_1)]$ for squares
compared with the Monte Carlo simulation energies $E_{0,1}$.

\vspace{1cm}

Fig.6. The gluon field overlap factors $f(E_{0,1})$ and $\bar{f}$(dashed line),
 as defined by eqs.(\ref{f's},\ref{dmin}), for a series of rectangles of
size $(d/a,r/a)$.

\vspace{1.0cm}

Fig. 7.  The values of  $k(E_0),k(E_1)$ and $\bar{k}$ corresponding  to the
$f(E_0),f(E_1)$ and $\bar{f}$ from tables 3 and 4 -- squares only, so that
$\alpha=4k/3$.

\vspace{1.0cm}

Fig. 8. A comparison of $k_0$ with $3\alpha_0/4$ and  $\bar{k}$ with
$3\bar{\alpha}/4$ .


\begin{thebibliography}{99}
\bibitem{som} R.Sommer and J.Wosiek, Nucl.Phys. {\bf B267} (1986) 531
\bibitem{Markum}
H. Markum, M.Meinhart, G.Eder, M.Faber and H.Leeb, Phys.Rev.{\bf D31} (1985)
2029
\bibitem{Ohta}
S.Ohta, M.Fukugita and A.Ukawa, Phys.Lett. {\bf B173} (1986)  15
\bibitem{CM1}
S.Perantonis, A.Huntley and C.Michael, Nucl.Phys. {\bf B326} (1989) 544
\bibitem{CM2}
S.Perantonis and C.Michael, Nucl.Phys. {\bf B347} (1990) 854
\bibitem{GMP}
A.M.Green, C.Michael and J. Paton,  Phys.Letts.{\bf B280}(1992)11
\bibitem{Hunt}
A.Huntley and C.Michael, Nucl.Phys. {\bf 270} (1986) 123

A.Huntley, Ph.D. thesis "Potentials from Lattice Gauge Theory" --
University of Liverpool (1988).
\bibitem{Rebbi}
C.B.Lang and C.Rebbi, Phys.Letts. {\bf B115} (1983) 351
\bibitem{chrism}
C.Michael, Nucl.Phys.B (Proc.Suppl.) {\bf 17} (1990) 59
\bibitem{Booth}
S.P.Booth et al., Phys.Letts. {\bf B275} (1992) 424
\bibitem{Gut}
 F.Gutbrod, Zeit.Phys. {\bf C30} (1986) 585
\bibitem{Paton}
A.M.Green and J.E.Paton, Nucl.Phys. {\bf A492}(1989)595
\bibitem{lenz}
F.Lenz et al. , Ann. Phys. (N.Y.) {\bf 170} (1986) 65

K.Masutani, Nucl.Phys. {\bf A468} (1987) 593

\bibitem{Masud}
B.Masud, J.E.Paton, A.M.Green and G.Q.Liu, Nucl. Phys. {\bf A528} (1991) 477
\bibitem{MGP}
O.Morimatsu, A.M.Green and J.E.Paton, Phys. Lett. {\bf B258} (1991)  257
\bibitem{Morimatsu}
O.Morimatsu, Nucl.Phys. {\bf A505} (1989)  655
\bibitem{Morimatsu2}
C.Alexandrou, T.Karapiperis and O.Morimatsu, Nucl.Phys. {\bf A518} (1990)
723
\bibitem{GMPS}
A.M.Green, C.Michael, J.E. Paton and M.E.Sainio, in preparation.
\end{thebibliography}
\end{document}